\documentclass[11pt,a4paper]{article}
\usepackage[margin=2.5cm]{geometry}
\usepackage{amsmath,amssymb}
\usepackage{graphicx}
\graphicspath{{./}{../github/figures/}}
\usepackage{hyperref}
\usepackage{booktabs}
\usepackage{authblk}
\usepackage{abstract}
\usepackage{titlesec}
\usepackage{natbib}
\usepackage{xcolor}

\titleformat{\section}{\large\bfseries}{\thesection.}{0.5em}{}
\titleformat{\subsection}{\normalsize\bfseries}{\thesubsection.}{0.5em}{}

\hypersetup{colorlinks=true,linkcolor=blue,citecolor=blue,urlcolor=blue}

\title{\textbf{The Triadic Stress Index in Financial Markets}}

\author[1]{Alberto Acedo}
\affil[1]{Biome Makers Inc. \quad \texttt{acedo@biomemakers.com}}
\date{August 2026}

\begin{document}
\maketitle

\begin{abstract}
Recent surveys of soil microbial networks under contrasting management report a
consistent picture: agricultural inputs raise modularity and shift network
organisation toward compartmentalised configurations that their authors read as
less mutualistic and more fragmented~\cite{saati2026}. The same literature
proposes network properties themselves, rather than species counts, as
biomarkers of soil condition~\cite{ortiz2021}. What it does not supply is a
mathematical quantity for the thing being described, which is the emergence of
\textbf{structural monopoly}: a few nodes accumulating the interactions the rest
lose. The accompanying theoretical work constructs that quantity, from
topological entropy, Wasserstein curvature bounds and the non-equilibrium
thermodynamics of complex networks~\cite{acedo2026fsri}. This paper takes that
framework, without alteration, to the correlation network of financial assets,
and asks what it says about systemic stress.

We call the resulting quantity the \textbf{Triadic Stress Index}, TSI. The
contribution is a conjunction rather than a single number: node-level
attribution included with no parameter to get wrong, detection on a par with the
best spectral measure, the cleanest alarms of anything tested, and a different
underlying primitive. The rest of this abstract develops each in that order.

TSI is built from the number of closed triangles in the weighted
network, together with density, modularity and degree variance. Its orientation is reversed for this domain, because
the configuration that means health in a soil community means crisis in a
market. We test it across \textbf{five markets}, equities including banking
crises and the AI sector, cryptocurrencies, commodities, foreign exchange and
sovereign debt, spanning 2006--2026, against labelled crisis episodes, at a
fixed alarm budget, out of sample, with block-bootstrap intervals.

\textbf{What is different about it comes first.} The index carries a per-node
decomposition, $\mathrm{diag}(A^3)$, that names which asset is carrying the
concentration, and it has no parameter to select. Spectral attribution must
choose how many components to read: with the right rule it matches the triangle
count, with a fixed $k=2$ it loses a quarter of its accuracy when that value is
wrong, and with the equally standard Kaiser rule it collapses to zero. Against
the only published per-node alternative on correlation networks, the local
balance index, the triangle decomposition scores 0.97--0.99 against 0.33--0.84,
and the ordering survives a benchmark rebuilt to favour the rival. We also
report, because it is true and useful, that on real matrices the far simpler
node degree gives the same answer.

\textbf{On detection it matches the best available measure.} Under a
like-for-like protocol, TSI ties the effective rank and the Vendi score,
the sharpest spectral baselines we could find, and does so across two samples
and three independent crisis definitions. It beats the Absorption Ratio, the
industry standard used by MSCI and central banks, by a wide margin (F1 gap
0.273 out of sample, 95\% CI $[0.10,\,0.39]$, $p<0.0005$), although that margin
narrows under the strictest labelling. Its alarms are the cleanest of anything
tested, 4.0\% unexplained against 14.7\% and roughly 59\%, and it shares only
half of them with its nearest relative, so it is reading something adjacent
rather than identical.

\textbf{We are explicit about register.} A lead--lag analysis shows the
cross-correlation with the stress index peaking at zero lag and decaying
symmetrically: this is a coincident state index, not a forecast, and every
claim here is framed accordingly. Two families of event fall outside it by
construction, single-asset divergence and slow geopolitical compounding, and we
close the obvious remedy for the first with a null: signing the correlation
leaves the triangle count unchanged, because positive semi-definiteness bounds
how frustrated a triangle can be.

\vspace{0.5em}
\noindent\textbf{Code and data:} \url{https://github.com/BiomeMakers/TSI-OmegaS}
\end{abstract}

\tableofcontents
\vspace{1em}


\section{Introduction}
\label{sec:intro}

TSI began as a descriptor of microbial co-occurrence networks in
agricultural soils, where resilience indices were derived from network
transitivity and competitive exclusion~\cite{acedo2022indices}, and where the
same four quantities, clustering, density, modularity and the proportion of
exclusion edges, were observed to move together along a management gradient.
The accompanying theoretical work~\cite{acedo2026fsri} formalises that
construction and establishes that its central quantity, the trace of the cube
of the adjacency matrix, proportional to the number of closed triangles in the
graph, is domain-agnostic: any system representable as a weighted network
admits the same computation.

This motivates a natural question. If the four factors that separate a
managed soil community from an undisturbed one are computed instead on the
correlation network among financial assets, do they register the kind of
structural concentration that characterises a system under systemic stress,
the configuration Minsky described as fragility built up in plain
sight~\cite{minsky1992}? We call the resulting quantity the \textbf{Triadic
Stress Index}, TSI, and write it that way throughout, in prose and in
equations.

A note on which factor is active, because that does not travel between
constructions and it is the first thing a reader should check. The
theoretical work~\cite{acedo2026fsri} reports a direct measurement in a
setting where the adjacency matrix is obtained by passing weight correlations
through a bounded sigmoidal map. There the clustering factor $C$ is driven
against a constant and is numerically inert, and the operative factor is the
degree-variance term $\mathrm{Coex}$. That is a property of how $A$ is built,
not of the index. Here $A$ is a correlation matrix among assets, built without
any saturating nonlinearity, so $C$ is free to vary and contributes to
TSI as defined; we checked this rather than assumed it, and the excess
ratio $C/D$ departs from unity on these matrices, whereas under a sigmoidal
construction it equals unity to four decimal places. The two settings are
therefore consistent: the same four factors, with the clustering channel inert
in one construction and active in the other.

We are deliberate about the verb. This paper tests whether TSI
\emph{registers} such a state, not whether it \emph{forecasts} its onset.
Section~\ref{sec:leadlag} reports a lead--lag analysis showing the distinction
matters: TSI is a coincident state index, not a leading indicator, and we
frame every claim accordingly.

Three claims are made and each is stated so that it can fail.
\textbf{H1, detection}: the index separates labelled crisis windows from calm
ones at least as well as the sharpest spectral measures available, at an equal
alarm budget. It fails if a baseline beats it with an interval excluding zero.
\textbf{H2, attribution}: the per-node decomposition names the asset carrying
the concentration without any parameter to select. It fails if a rule with a
parameter beats it, or if a published per-node alternative does.
\textbf{H3, register}: the index is coincident with stress rather than leading
it. It fails if the cross-correlation with an external stress measure peaks at
a negative lag.

The order of what follows is deliberate: first what the index does that the
alternatives do not, then where it merely equals them, then the boundary of what
it can be asked to do at all. Results that did not survive testing are reported
alongside those that did, including the scope boundaries and the questions that
remain open.


\section{Hypotheses and What Would Falsify Them}

The three claims of Section~\ref{sec:intro} are operationalised as follows, and
the criteria were fixed before the corresponding experiments were run.

\begin{itemize}
\item \textbf{H1, detection at parity.} Scored by $F_1$ at the 90th percentile
  of each series against labelled crisis windows, so that every metric spends
  the same alarm budget, with block-bootstrap intervals on the difference.
  Falsified if any baseline's advantage has an interval excluding zero. Three
  independent crisis definitions are used and reported separately, because a
  result that holds only under one labelling is not a result.
\item \textbf{H2, attribution without a parameter.} Scored by the hit rate in
  the top-$|$epicentres$|$ on a synthetic benchmark where ground truth exists,
  across regimes with one to four simultaneous epicentres. Falsified if a
  spectral rule with a well-chosen $k$, or a published per-node alternative,
  scores higher across regimes.
\item \textbf{H3, coincident register.} Scored by the cross-correlation between
  the index and an external stress index across lags. Falsified if the peak
  sits at a negative lag, which would make it a leading indicator.
\end{itemize}

A fourth question is not a hypothesis but a scope condition, and it is answered
in Section~\ref{sec:large_network}: on what kind of network does the raw index
work, and where does it need the persistence filter?

\section{The Index}

\subsection{Construction on a Financial Network}

For a rolling window of $W$ days of daily log-returns of $n$ assets, we
build the correlation matrix $\rho$, define the adjacency matrix
$A = |\rho|$ (absolute value, zero diagonal), and compute:
\begin{equation}
\mathrm{TSI} = \frac{C \cdot D}{M} \cdot \mathrm{Coex},
\label{eq:omega}
\end{equation}
where $C$ is the weighted clustering coefficient (normalised
$\operatorname{Tr}(A^3)$), $D$ is connection density, $M$ is the inverse
spectral gap of the graph Laplacian (a modularity proxy), and
$\mathrm{Coex}$ is the variance of the degree sequence. The construction
descends from prior work on co-inclusion and co-exclusion networks in
agricultural microbiomes, where resilience indices were derived from network
transitivity~\cite{acedo2022indices}, following the
original definition in~\cite{acedo2026fsri}. Because $C$, $D$, $M$, and
$\mathrm{Coex}$ all scale with $n$ in different ways, raw TSI
magnitudes are \textbf{not comparable across networks of different size};
only within-series dynamics (z-scores, percentiles) are.

\paragraph{The name.} We call the quantity in~\eqref{eq:omega} the
\textbf{Triadic Stress Index} (TSI). Each part of the name records something
that the rest of this paper either establishes or bounds. \emph{Triadic},
because the closed-triangle count is the only one of the four factors that is
not a function of the degree sequence alone, and because the per-node
attribution of Section~\ref{sec:attrib} is the triangle count read per node
rather than summed. \emph{Stress}, because in this domain the index rises
under stress, which is the reverse of the orientation the same construction
carries in the ecological setting it comes from, for the reason set out
immediately below. And \emph{index} rather than signal or forecast, because
Section~\ref{sec:leadlag} establishes that it is coincident with stress and
does not lead it.

\paragraph{A note on notation, because the names in this family are easy to
confuse.} The accompanying theoretical work writes this quantity $\Omega$, and
the machine-learning line built on the same trace carries the Omega name as
well. To keep the financial instrument distinct from both, we write TSI
throughout this paper, in prose and in equations, and reserve the other names
for what they denote: \textbf{FSRI} for the theoretical framework,
\textbf{Omega-S} for the machine-learning line, and \textbf{TSI} for the index
defined here.

\paragraph{Relation to the composition fixed in the framework paper.}
\label{rem:composition_fin}
Equation~\eqref{eq:omega} places $\mathrm{Coex}$ in the numerator, whereas
Definition~1 of the framework paper~\cite{acedo2026fsri} places it in the
denominator, so that a high degree variance lowers the index there and raises it
here. The two agree on the other three factors, including the orientation of $M$
as an inverse spectral gap. We keep the composition above, and we give the
reason rather than leaving the difference to be found by comparing code.

The reason is that the two domains do not agree on what the same structure
means. Under stress, correlations rise, the network becomes denser and more
connected, and every factor moves in the same direction: $C$ rises, $D$ rises,
$\mathrm{Coex}$ rises, and $M$, being an inverse spectral gap, falls. A
stress indicator should therefore place the three that rise in the numerator and
the one that falls in the denominator, which is what~\eqref{eq:omega} does; the
composition of Definition~1 places $\mathrm{Coex}$ against the other three. The
underlying asymmetry is that a densely interconnected, weakly modular community
is the healthy configuration in the ecological setting the index derives from,
and is the crisis configuration in a market, where it means that everything
falls together. The difference between the two papers is thus a deliberate
reorientation by domain and not an unreconciled discrepancy.

We also tested it rather than only arguing it, on the same windows and with the
same detection protocol used throughout this paper. Over a portfolio of large
US banks from 2006 to 2011, with the crisis window labelled as 2007-08 to
2009-06, the filtered $F_1$ at the ninetieth percentile is $0.275$ for the
composition of~\eqref{eq:omega} and $0.157$ for that of Definition~1, against a
chance level of $0.152$ for an alarm rate of ten percent at the observed crisis
frequency. The composition used here detects at roughly $1.8\times$ chance while
the alternative sits at chance. Two limits should be read alongside that.
Block-bootstrap confidence intervals on the paired difference are wide and
include zero ($+0.054$, $[-0.229, +0.315]$), so the comparison is directional
rather than conclusive; and on this particular portfolio and period the
Absorption Ratio also sits near chance ($0.167$), which indicates that the test
itself discriminates weakly rather than that the alternative composition is
uniquely poor. The argument from domain orientation above is the stronger of the
two and does not depend on this measurement.

A third composition deserves reporting because its outcome is a limitation of
the index rather than a point in its favour. The framework paper shows that
$\mathrm{Coex} = \operatorname{Var}(k)$ carries units of degree squared and
proposes a shape-normalised variant that replaces it by the Gini coefficient of
the degree sequence and the density by the mean degree, removing the dependence
on the scale of the degree sequence. On the same test that variant scores
$0.098$, that is \emph{below} chance. The implication is uncomfortable and we
state it: a substantial part of what TSI detects in this setting is the
level of correlation rather than the shape of its distribution, and removing the
scale dependence removes the signal with it. This does not affect the
comparisons reported below, which are all against external baselines computed on
the same windows, but it does bound what the index should be claimed to measure.

\paragraph{The closest structural precedent, and we do benchmark against it.}
Sandhu, Georgiou and Tannenbaum~\cite{sandhu2016} compute Ollivier-Ricci
curvature on correlation networks of equities and report it as an indicator of
market fragility and systemic risk. That is the same graph construction and the
same purpose as the present work, reached from discrete geometry rather than
from triadic structure, and it is the nearest precedent to what follows.
Earlier versions of this paper named it and declined to benchmark against it,
on the grounds that per-edge Ollivier-Ricci requires a Wasserstein computation
for every edge and therefore sits in a different cost regime from a stochastic
trace estimate. That was a reason not to have done the work, not a reason for
it not to be done, so we have now done it: Section~\ref{sec:ricci} reports the
head-to-head on the same windows, under the same filter and the same alarm
budget as every other benchmark here. We note in passing that a subsequent
analysis~\cite{expost2024} argues the curvature identifies the size and
duration of a crisis accurately without leading it, which is the same register
we establish for TSI in Section~\ref{sec:leadlag}.

\paragraph{Other network-structural indicators, and where this sits.} Systemic
risk has been approached through exposure and connectedness
models~\cite{diebold2014,acemoglu2015,billio2012} and through random-matrix
cleaning of correlation matrices~\cite{laloux1999,plerou2002}. The present
work belongs to neither: it fixes a single composite index in advance, taken
from an external domain, and asks how it scores against spectral baselines on
the same windows. Three lines of work are close enough that we state the
relation explicitly rather than let a reader find it. Samal, Kumar, Yadav and
Chakraborti~\cite{samal2021fin} evaluate a battery of network-centric
indicators, including edge curvatures, clique number and community counts, on
threshold networks over minimum spanning trees of 69 global indices, and
conclude that network measures are useful for monitoring fragility. That is
the same question as ours; what differs is that they survey measures and we
commit to one in advance. Bartesaghi, Diaz-Diaz, Grassi and
Uberti~\cite{bartesaghi2025} use the global balance index of the
\emph{signed} correlation network as a systemic-risk measure and derive a
local, per-node version of it, developed further into an investment-decision
rule in follow-up work~\cite{bartesaghi2025local}. That is the closest existing
work to the signed extension this paper leaves open, and we state the position
plainly: our $A = |\rho|$ discards precisely the information their index is
built from, their global balance is already benchmarked against spectral risk
measures as ours is, and their local balance is a per-node attribution of the
same kind as ours. Rather than list the signed variant as future work, we ran
it, and report three findings in Section~\ref{sec:signed}. Zhang, Wang, Zheng and
Cartlidge~\cite{zhang2026defi} pair a correlation fragility indicator computed
on time-varying correlation networks with a risk contribution score that
attributes systemic risk to categories of protocol; a structural index plus a
per-node decomposition is the same shape as what is reported here, on a
different asset universe. None of the three is used as a running benchmark,
for the cost-regime reason given above, and we regard a comparison against the
signed balance index as the most informative of the three left undone.

\section{Benchmarks}
\label{sec:benchmarks}

Four families of comparison are used, and the reason for each differs.

\begin{itemize}
\item \textbf{The Absorption Ratio}, described below. It is the industry
  standard and therefore the measure a practitioner would already have running.
  It is a weak baseline on this data, which the results below make plain.
\item \textbf{The effective rank and the Vendi score}, spectral entropies of
  the correlation eigenvalues. These are the \emph{sharpest} baselines we could
  find rather than the most common ones, and they are the honest test of H1.
  They are described and scored in Section~\ref{sec:spectral}.
\item \textbf{Per-node alternatives}: spectral attribution under four different
  rules for choosing the number of components, the node degree, and the local
  balance index of Bartesaghi et al. All are scored in
  Section~\ref{sec:attrib}.
\item \textbf{The global balance index} of the signed correlation network,
  treated separately below because it cannot be scored fairly on the networks
  used here.
\end{itemize}

\subsection{Benchmark: the Absorption Ratio}
\label{sec:ar}

The Absorption Ratio~\cite{kritzman2010} is the fraction of total variance
explained by the first $k$ principal components of the return covariance
matrix ($k = \lfloor n/5 \rfloor$). It is the de facto standard in
systemic-risk management, used by MSCI and central banks, and is computed
on the same rolling windows as TSI for direct comparability.

\textbf{Additional spectral benchmarks.} The Absorption Ratio summarises the
eigenvalue spectrum through the mass of its leading components. Because
TSI is itself partly a spectral object, benchmarking against the AR alone
risks comparing against a deliberately coarse summary. We therefore add two
sharper spectral baselines on the same windows: the \emph{effective
rank}~\cite{roy2007}, $\exp(H)$ with $H$ the Shannon entropy of the
correlation eigenvalues normalised to sum to one; and the \emph{Vendi
score}~\cite{friedman2023}, the exponential of the
von Neumann entropy of the normalised correlation kernel, which on a
correlation matrix is \emph{mathematically identical} to the effective rank,
an equivalence we state explicitly rather than presenting the two as
independent evidence. We additionally report an effective rank computed on the
return matrix itself (entropy of the singular-value spectrum), a genuinely
distinct quantity.

\subsection{The Global Balance Index, and Why It Is Not in the Main Table}
\label{sec:balancebench}

The nearest published rival is the global balance index of the signed
correlation network~\cite{bartesaghi2025}, $\kappa = \operatorname{tr}(e^A) /
\operatorname{tr}(e^{|A|})$, with a per-node counterpart used for investment
decisions in follow-up work~\cite{bartesaghi2025local}. It belongs in this list
on merit: it is built on the same object, aims at the same target, and is
published in a peer-reviewed venue. We reimplemented it and verified our
version against the authors' own public code, obtaining agreement to machine
precision over 2\,232 windows.

It does not appear as a column in the detection table for a reason we measured
rather than assumed. On small networks the index \emph{saturates}: with eight
nodes it sits at its maximum of 1 on 26\% of windows in its weighted form and
70\% in the binary form the same authors report, because a small correlation
network is close to balanced whatever the market is doing. Any ranking produced
under those conditions describes a degenerate regime, not the index, so we
produce none. Section~\ref{sec:signed} reports what we did measure instead,
including the regime in which the index is well behaved, a check that its own
published relationship reproduces on our data, and a head-to-head on a
464-asset universe where it does not saturate.

\section{Data and Protocol}

\subsection{Data}

\begin{table}[ht]
\centering
\caption{Datasets used in this study.}
\label{tab:data}
\begin{tabular}{llll}
\toprule
Scenario & Assets ($n$) & Period & Source \\
\midrule
Banking crises (Bear Stearns, Lehman, & 10 banks & 2006--2011 & yfinance \\
European debt crisis) & & & \\
AI sector & 11 stocks & 2021--2026 & yfinance \\
Out-of-sample validation & 8 OFR components & 2000--2026 & OFR \\
Large-network robustness & 42 stocks, 7 sectors & 2006--2026 & yfinance \\
Cross-market (crypto, FX, commodities, & 10+7+10+10 & 2021--2024 & yfinance \\
sovereign debt) & & & \\
\bottomrule
\end{tabular}
\end{table}

All windows: 20 trading days, recomputed every 3 days.


\subsection{TSI with Persistence Memory}
\label{sec:mem}

We apply an asymmetric exponential filter:
\begin{align}
\mathrm{mem}(t) &= \alpha_{\uparrow}\,\mathrm{TSI}(t) +
  (1-\alpha_{\uparrow})\,\mathrm{mem}(t-1)
  \quad \text{if } \mathrm{TSI}(t) > \mathrm{mem}(t-1), \label{eq:mem_up}\\
\mathrm{mem}(t) &= \alpha_{\downarrow}\,\mathrm{TSI}(t) +
  (1-\alpha_{\downarrow})\,\mathrm{mem}(t-1)
  \quad \text{if } \mathrm{TSI}(t) \leq \mathrm{mem}(t-1), \label{eq:mem_down}
\end{align}
with $\alpha_{\uparrow} = 0.6$ and $\alpha_{\downarrow} = 0.08$, chosen
by inspection and later validated by grid search (Section~\ref{sec:oos}).

\subsection{Out-of-Sample Calibration and Significance Testing}
\label{sec:oos}

Hyperparameters were calibrated via grid search
($\alpha_{\uparrow} \in [0.3, 0.8]$, $\alpha_{\downarrow} \in [0.02, 0.20]$)
with a genuine temporal train/test split on the 2000--2026 OFR Financial
Stress Index series (train 2000--2015, test 2016--2026). Statistical
significance of the F1 gap between TSI with memory and the Absorption
Ratio was assessed via block bootstrap (block length $\approx$ 1 quarter,
$B = 2{,}000$ resamples), restricted to the out-of-sample period only.

\subsection{Persistent Homology}

We implemented Vietoris-Rips persistent homology from scratch
(boundary-matrix reduction, $\mathbb{Z}_2$ coefficients), validated against
an analytical test case (a 6-point circle, which yields exactly one H1
generator). We tested two constructions:

\textbf{Correlation-network H1} (negative result): Vietoris-Rips filtration
on the asset correlation-distance matrix $d_{ij} = \sqrt{2(1-\rho_{ij})}$
\cite{mantegna1999}.

\textbf{Takens-embedding H1} (positive result): following Gidea \&
Katz~\cite{gidea2018}, a scalar stress series is reconstructed as a point
cloud in $\mathbb{R}^d$ via time-delay embedding, and persistent homology
is computed on that point cloud within each rolling window.

\section{Results}

What follows is ordered by argument rather than by chronology: first the
layer that has no equivalent among the baselines, then the layer where the
index merely matches them, then the comparison it wins and the one it does not,
then the boundary of what it can be asked to do, and finally the robustness
checks and the extension we closed with a null.

\subsection{On the Node-Attribution Layer}
\label{sec:attrib}

Detection says the system is stressed. Attribution says \emph{where}. The two
are not equally hard, and the difference is not that spectral methods cannot
attribute, because they can: the loading on the leading eigenvector, directly
available to any Absorption-Ratio user, recovers a single epicentre perfectly.
The difference is that they need to be told how many places to look, or given a
rule for deciding it. Reading eigenvectors requires committing to $k$, the
number of components to inspect, and in deployment $k$ is either fixed in
advance, before anyone knows how many epicentres the next episode will have, or
selected per window by a rule that can itself be wrong. The per-node
decomposition $\mathrm{diag}(A^3)$, which is the alarm itself read per node
rather than summed, has no such parameter. How much that is worth is the
question this section answers, and the answer is more modest than the framing
above suggests.

We tested this on synthetic networks with a known epicentre, in two steps.
First, whether the attribution works at all: over 80 simulations the culprit
rule recovers the true epicentre in essentially every run (top-3 hit rate 1.00
against a chance level of 0.25), and under two simultaneous uncorrelated
epicentres it recovers 0.93 of the six affected nodes (chance 0.50). Then the
question that decides the comparison, which is robustness when the number of
epicentres is unknown. We ran regimes of one to four simultaneous epicentres,
holding the spectral method at a fixed $k=2$ as it would be held in deployment
(120 simulations per regime, paired Wilcoxon):

\begin{center}
\begin{tabular}{lcccccc}
\toprule
 & Triangles & & \multicolumn{4}{c}{Spectral, by how $k$ is chosen} \\
\cmidrule(lr){5-7}
Epicentres & $\mathrm{diag}(A^3)$ & Degree & fixed $k=2$ & MP & Kaiser & oracle \\
\midrule
1 & \textbf{0.994} & 0.933 & 0.756 & 0.997 & 0.000 & 0.989 \\
2 & 0.978 & 0.940 & 0.999 & 0.978 & 0.071 & 0.999 \\
3 & \textbf{0.980} & 0.952 & 0.946 & 0.962 & 0.457 & 0.981 \\
4 & \textbf{0.975} & 0.961 & 0.948 & 0.959 & 0.808 & 0.989 \\
\bottomrule
\end{tabular}
\end{center}

\noindent Top-$|\text{epicentres}|$ hit rate, 120 simulations per regime, 16
nodes and 40-observation windows. \emph{MP} selects $k$ per window as the
number of eigenvalues above the Marchenko-Pastur edge~\cite{marchenko1967}
$(1+\sqrt{n/T})^2 = 2.665$, the standard random-matrix rule; it picks
$k = 1.71$ on average. \emph{Kaiser} selects the eigenvalues above 1 and picks
$k = 5.87$. \emph{Oracle} is given the true number of epicentres and is not
available in practice; it is the ceiling.

\begin{figure}[ht]
\centering
\includegraphics[width=0.72\textwidth]{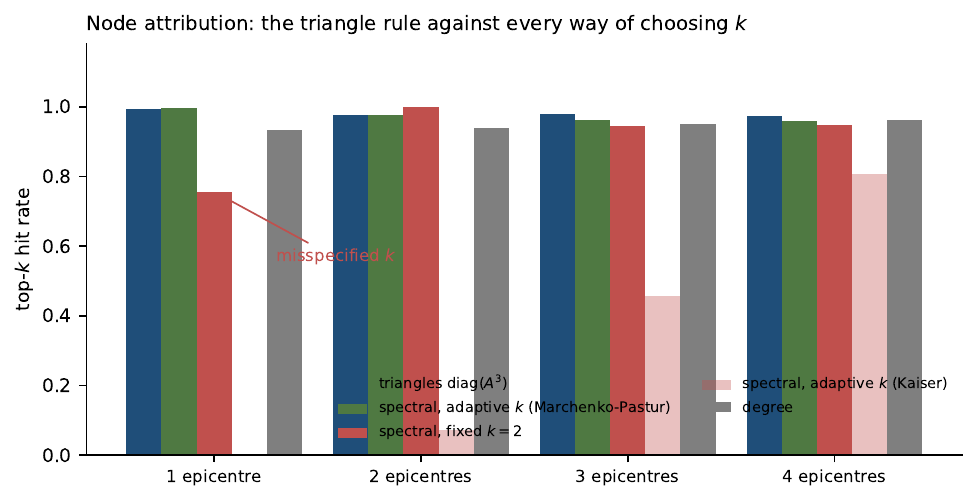}
\caption{Attribution accuracy as the number of simultaneous epicentres varies.
The triangle rule is flat across regimes because it has no parameter to
misspecify. The spectral rule depends entirely on how $k$ is chosen: the
Marchenko-Pastur rule ties the triangle rule everywhere, a fixed $k=2$ costs a
quarter of the accuracy when it is wrong, and the equally standard Kaiser rule
collapses. The spread between those three, not the gap to the triangle rule, is
the result.}
\label{fig:attribution}
\end{figure}

The triangle rule is flat, 0.975 to 0.994 across every regime, because there is
no parameter in it to misspecify. Fixed $k=2$ matches it exactly where that
value happens to be right (two epicentres) and loses a quarter of its accuracy
where it is not (0.756 with a single epicentre).

\textbf{The adaptive-$k$ control, and what it costs the claim.} An earlier
version of this section stopped there. It should not have, because the obvious
objection is that a practitioner would not hold $k$ fixed but read it off the
spectrum window by window, and we had not tested it. We have now, and the
result is against us: with $k$ chosen by the Marchenko-Pastur edge the spectral
rule scores 0.959 to 0.997 and ties the triangle rule across all four regimes.
The oracle ties it as well. The gap in the fixed-$k$ column is therefore not
evidence that triangles see something eigenvectors cannot; it is evidence that
$k$ matters, and a good rule for choosing $k$ closes it.

What the control does not do is make the choice free. The three selection rules
in the table span 0.000 to 0.997 on the single-epicentre regime: Marchenko-Pastur
lands near the ceiling, a fixed $k=2$ costs a quarter of the accuracy, and the
Kaiser rule ($\lambda > 1$), which is equally standard and equally defensible a
priori, collapses completely, because it selects roughly six eigenvectors and
the epicentre stops being separable in their combined loading. The spectral
route can match $\mathrm{diag}(A^3)$, but only conditional on a selection
decision whose wrong answers are catastrophic and are not identifiable without
the ground truth we have here and a practitioner does not.

So we state the claim at the size the evidence supports, which is smaller than
the one we set out to make. TSI's attribution is not more accurate than
spectral attribution; against a well-chosen adaptive rule it is exactly as
accurate. What it is, is unconditional: it arrives with the index, ranks the
nodes with no parameter to select and no way to select it wrongly, and it also
beats the simpler degree rule in all four regimes (0.933--0.961), which is why
we adopt it over a generic centrality. That is a robustness and simplicity
argument, not a resolution argument, and it should be read as one.

Two further limits bound even that. The evidence is synthetic, with a
controlled ground truth real markets do not provide, so it is a mechanism
result and not a field validation. And it does not overturn the detection tie
of Section~\ref{sec:spectral}: on discriminating stressed windows from calm
ones, TSI and the effective rank remain statistically indistinguishable,
and nothing here changes that.

\subsection{Benchmarking Against Sharper Spectral Baselines}
\label{sec:spectral}

Because the Absorption Ratio is a deliberately coarse spectral summary, we
re-ran the identical protocol (same windows, crisis labels, F1-at-90th-percentile
criterion and block bootstrap) against the sharper baselines of
Section~\ref{sec:ar}. One design point decides this comparison and we handle it
explicitly: the persistence filter of Section~\ref{sec:mem} helps any series
that is elevated in bursts, so filtering TSI and not its rivals would
flatter TSI. We therefore report every metric twice, raw and under the
same asymmetric filter, and run the bootstrap on the like-for-like column.
Out of sample (2016--2026, 897 windows):

\begin{center}
\begin{tabular}{lcccccc}
\toprule
 & \multicolumn{2}{c}{F1@p90} & & & & \\
\cmidrule(lr){2-3}
Metric & raw & filtered & $\Delta$F1 & 95\% CI & $p$ & Verdict \\
\midrule
TSI & 0.367 & \textbf{0.447} & \textendash & \textendash & \textendash & \textendash \\
Effective rank (corr.) & 0.347 & 0.377 & $+0.038$ & $[-0.010, +0.111]$ & 0.10 & tie \\
Vendi score & 0.347 & 0.377 & $+0.037$ & $[-0.009, +0.104]$ & 0.12 & tie \\
Absorption Ratio & 0.174 & 0.134 & $+0.219$ & $[+0.070, +0.410]$ & $<0.0005$ & TSI wins \\
Effective rank (returns) & 0.179 & 0.174 & $+0.227$ & $[+0.097, +0.393]$ & $<0.0005$ & TSI wins \\
\bottomrule
\end{tabular}
\end{center}

\noindent$\Delta$F1 and its interval come from the block bootstrap on the
filtered column and need not equal the difference of the two point estimates.
The filter is the same asymmetric map for every row, applied to the
stress-oriented signal (the effective rank and the Vendi score fall under
stress, so they enter negated).

\begin{figure}[ht]
\centering
\includegraphics[width=0.72\textwidth]{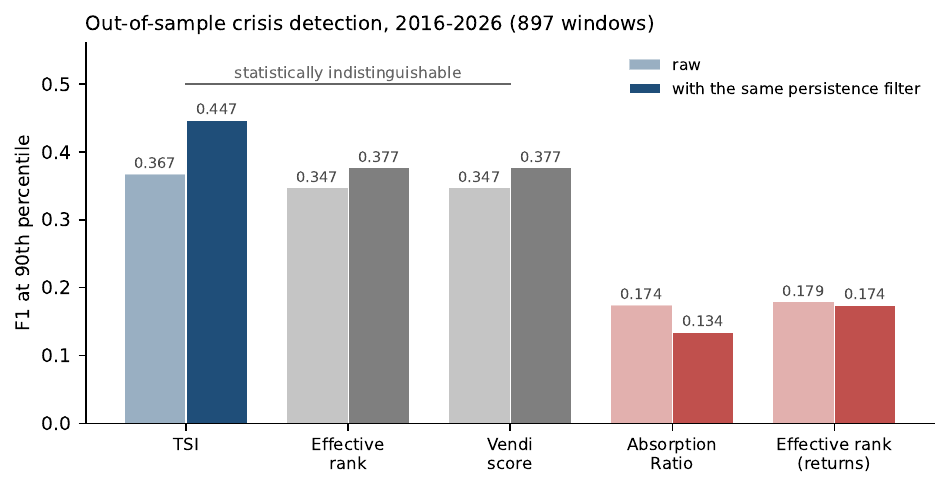}
\caption{Out-of-sample crisis detection, each metric read raw (pale) and under
the same persistence filter (solid). TSI outperforms the Absorption Ratio
by a wide margin in both readings, but the sharper spectral baselines sit
within its confidence interval in both. The advantage over the industry
standard is real; the advantage over a properly sharpened spectral measure is
not, and giving the rivals the same filter does not change that verdict.}
\label{fig:benchmarks}
\end{figure}

The verdict does not depend on which column is read. Raw, TSI leads the
effective rank by 0.020; filtered, by 0.070; in neither case does the bootstrap
interval exclude zero. Over the full history (2000--2026, 2{,}232 windows) the
pattern repeats: under the same filter TSI scores 0.389 against 0.342 for
the effective rank (tie, $p = 0.07$) and 0.175 for the Absorption Ratio
(TSI wins, $p < 0.0005$).

\textbf{Does the verdict depend on the crisis list?} For one comparison, yes.
Re-running the same protocol under each of the three lists of
Section~\ref{sec:oos_results}, on both the hold-out period and the full
history, with every series filtered:

\begin{center}
\begin{tabular}{llcccc}
\toprule
Sample & List & TSI & Eff.\ rank & Absorption R. & Eff.\ rank (ret.) \\
\midrule
Out of sample & 16 ep.\ (21.9\%) & 0.434 & 0.343 $=$ & 0.168 $=$ & 0.224 $\bullet$ \\
2016--2026    & 23 win.\ (34.9\%) & 0.447 & 0.377 $=$ & 0.134 $\bullet$ & 0.174 $\bullet$ \\
              & 25 ep.\ (33.0\%) & 0.451 & 0.383 $=$ & 0.135 $\bullet$ & 0.181 $\bullet$ \\
\midrule
Full history  & 16 ep.\ (27.7\%) & 0.340 & 0.363 $=$ & 0.204 $=$ & 0.192 $=$ \\
2000--2026    & 23 win.\ (40.6\%) & 0.389 & 0.342 $=$ & 0.175 $\bullet$ & 0.165 $\bullet$ \\
              & 25 ep.\ (39.8\%) & 0.386 & 0.343 $=$ & 0.164 $\bullet$ & 0.167 $\bullet$ \\
\bottomrule
\end{tabular}
\end{center}

\noindent$=$ tie, $\bullet$ TSI wins, by block bootstrap on the F1 gap.
Percentages are the share of windows the list labels as stress.

Two readings follow. The tie against the effective rank is \emph{robust}: it
holds in all six cells. Its direction is consistent, with TSI ahead in
five of the six, and in the sixth the effective rank leads by 0.023, well
inside the noise ($p = 0.59$). The win over the Absorption Ratio is \emph{not}
robust: it disappears in both cells of the narrowest list, where the gap
remains positive ($+0.170$ and $+0.121$) but the interval touches zero. We
report this because the Absorption Ratio is the benchmark this work set out
to beat and the one a practitioner would already have in production. The
correct statement is that TSI beats it clearly under the two broader
labellings and falls short of conventional significance under the narrowest,
which is also the one with the fewest labelled windows and the least power.

\textbf{Why the tie happens, and what it is not.} The four baselines are not
four independent objects. The Vendi score and the effective rank are
identical by construction on a correlation matrix, as noted in
Section~\ref{sec:ar}, and empirically the Absorption Ratio and the effective
rank computed on the return matrix are near-duplicates as well (Pearson 0.91,
Spearman 0.97), which is why they score alike. That leaves two distinct
baselines, not four. Against them TSI is far from equidistant: it
correlates 0.85 (Spearman 0.93) with the effective rank and only 0.54
(Spearman 0.61) with the Absorption Ratio, and it shares 53\% of its alarm
windows with the former against 34\% with the latter. TSI is therefore a
close relative of the effective rank and a distant one of the Absorption
Ratio, which is exactly what the companion theoretical work predicts: the
global scalar $\operatorname{Tr}(A^3)$ is the third spectral moment and
belongs to the same family as the spectral entropies~\cite{acedo2026fsri},
whereas the Absorption Ratio reads leading-eigenvalue mass. The tie is not an
awkward null result to be explained away; it is the predicted outcome, and the
47\% of alarms TSI does not share with its nearest relative is the
measure of how far a different construction of the same spectrum can move.

\textbf{Alarm quality on a common criterion.} Section~\ref{sec:oos_results}
gave an unexplained-alarm rate for TSI alone. Computed for every metric on
the same 25-episode list, at the same top-decile alarm budget (full history,
all series filtered): TSI 4.0\% (9 of 224), effective rank and Vendi
14.7\% (33), effective rank on returns 58.5\% (131), Absorption Ratio 59.4\%
(133). Out of sample the ordering is the same (3.3\%, 17.8\%, 61.1\%, 71.1\%).
TSI's alarms are therefore the cleanest of any metric tested, by a wide
margin over the industry standard and by roughly a factor of four over the
effective rank.

One qualification prevents that from being read as an independent result, and
it is the reason we ran the comparison rather than quoting the 4\% on its own.
With the alarm budget fixed at the top decile, the unexplained-alarm rate is
exactly $1 - \text{precision}$, and F1 at a fixed alarm count is a monotone
function of the same quantity. The two comparisons are one comparison in two
presentations. The verdict from the bootstrap therefore carries over unchanged:
against the effective rank the difference in alarm quality is real in point
estimate and not separable from zero at this sample size, exactly as the F1
comparison found.

We read this as the honest and more informative result. TSI's advantage
over the Absorption Ratio is real, reproducible and statistically significant
but the gap is largely explained by the AR being a weak baseline on this
data (F1 $\approx 0.17$--$0.19$, roughly half of every other measure tested),
not by TSI accessing information the spectrum lacks. \textbf{Against a
properly sharpened spectral baseline, TSI ties.} It does not lose, and
retains a small positive point estimate in every comparison, but the difference
is not statistically distinguishable from zero. The defensible claim is
therefore not that TSI dominates spectral methods, but that it matches the
best of them while being constructed from a different primitive, network
topology rather than variance decomposition, and while carrying an attribution
layer that arrives with the index and needs no selection rule
(Section~\ref{sec:attrib}, which also reports what that is and is not worth). A lead--lag
analysis (Section~\ref{sec:leadlag}) establishes the correct register for all
of these claims: the cross-correlation with the stress index peaks at zero lag
and decays symmetrically, so TSI is a \emph{coincident state index},
reporting that the network is presently in a concentrated and stressed
configuration, and not a leading indicator. We frame every claim accordingly.

\subsection{Out-of-Sample Calibration and False-Alarm Rate}
\label{sec:oos_results}

Over the full 2000--2026 OFR series (2,232 windows), the correlation between
in-sample and out-of-sample F1 across the full 36-combination grid is
$\mathbf{0.96}$. The optimum ($\alpha_{\downarrow} = 0.08$, robust for
$\alpha_{\uparrow} \in [0.5, 0.7]$) matches the values chosen by inspection.

\textbf{Three ground-truth lists, and which figure uses which.} The episodes
that count as stress were labelled at three points in this project and the
lists are not identical. We set them out rather than let a reader assume a
single list throughout, because each headline number below is attached to a
different one.

\begin{itemize}
\item A \emph{16-episode} list with narrow windows, used for the
  hyperparameter calibration of Section~\ref{sec:oos}. On it, out of sample
  (2016--2026), TSI with memory gives \textbf{69\% precision and 32\%
  recall} at the 90th-percentile threshold.
\item A \emph{23-window} list with wider windows, used for every F1 figure in
  Sections~\ref{sec:signif} and~\ref{sec:spectral}. On it, out of sample, the
  same series gives precision 1.00 and recall 0.29 (90 alarms, none falling
  outside a labelled window).
\item A \emph{25-episode} list used to ask a different question: not whether
  an alarm falls inside a pre-drawn window, but whether it corresponds to any
  documented episode at all. Across the full history 9 of 224 alarms have no
  such correspondence, a \textbf{4.0\% unexplained-alarm rate}.
\end{itemize}

Three caveats bound that last figure and we state all three. It is computed
over the full 2000--2026 history rather than the hold-out period, so it is not
an out-of-sample result. Its list is broader than the 23-window one, so it is a
laxer test by construction. And we have not computed the same quantity for the
Absorption Ratio or for the spectral baselines, so it characterises TSI
and does not separate it from them. Rather than pick one list and discard the others, we re-ran every comparison
under all three. Section~\ref{sec:spectral} reports the result: one verdict in
this paper does depend on which list is used, and it is not the one we
expected.

\subsection{Statistical Significance}
\label{sec:signif}

Restricted to 2016--2026 (897 windows), the F1 gap between
TSI with memory and the Absorption Ratio is 0.273 (0.447 vs.\ 0.174).
Block-bootstrap ($B=2{,}000$, block $\approx$ 1 quarter):
\textbf{95\% CI $[0.095,\, 0.392]$}, does not cross zero, one-sided
$p < 0.0005$.

TSI with memory here is the asymmetric filter of Section~\ref{sec:mem}
($\alpha_{\uparrow} = 0.6$, $\alpha_{\downarrow} = 0.08$) applied to raw
TSI, which is the form used throughout this paper. We say so explicitly
because the next section compares against spectral rivals and the choice of
filter moves the number materially.

\emph{Caveat.} This metric rewards sustained-plateau behaviour, which is
precisely what the memory filter was built to produce. The test confirms the
filter achieves its design goal with statistical confidence, not that
TSI with memory is superior on some independent criterion.

\paragraph{The head-to-head, run where the index is well behaved.} On 464
S\&P~500 constituents with a complete ten-year history the balance index does
not saturate: it takes distinct values in every window, and at the 90th
percentile it flags 10.2\% of them, as do TSI, the binary balance index and
the effective rank. The Absorption Ratio cannot be scored in this
configuration at all, and for a reason worth stating: with a 60-day window and
464 assets the covariance matrix has rank 59, so the top $n/5 = 93$ eigenvalues
capture the entire trace by construction and the ratio is identically one. That
is a limitation of the configuration, not of the measure.

Ground truth here is \emph{theirs}, not ours: a window is labelled systemic when
the cross-sectional mean return falls below a threshold over a 20-day span,
which is the definition used in the original work. Adopting it removes our own
episode judgement from the comparison entirely.

\textbf{The result is a tie, at every threshold and under both readings of the
label.} Scoring against the contemporaneous span, TSI reaches $F_1$ of
$0.160$--$0.161$ and the balance index $0.184$--$0.187$, with every paired
difference's interval crossing zero ($p$ from $0.43$ to $0.63$); the binary
variant and the effective rank land in the same place. No measure separates
itself, and none is far above the chance level of roughly $0.14$ at this alarm
budget and base rate. We read that as a property of the label rather than of the
measures: a return-based event is a coarse proxy for a structural one, and this
is the price of using someone else's definition rather than our own.

\textbf{One thing did fall out of it, and it corroborates the central claim of
this paper from an unexpected direction.} Orientation was measured rather than
assumed in every cell. Against the \emph{contemporaneous} span TSI enters
positively, rising as returns deteriorate. Against the \emph{forward} span its
orientation reverses. An index that reads the present correctly and inverts when
asked about the future is precisely a coincident state index, which is what
Section~\ref{sec:leadlag} establishes by a different route. These tests use a
different universe and a different ground truth from Section~\ref{sec:spectral},
so they form their own family and are not folded into the correction of
Section~\ref{sec:multiplicity}.

\subsection{Head-to-Head Against Ollivier--Ricci Curvature}
\label{sec:ricci}

Correlations are mapped to the Mantegna distance $d_{ij} = \sqrt{2(1-\rho_{ij})}$,
each node carries a measure over its neighbours proportional to the absolute
correlation, and the curvature of an edge is
$\kappa_{ij} = 1 - W_1(m_i,m_j)/d_{ij}$ with the earth-mover distance computed
exactly by linear programming. The market-level scalar is the mean edge
curvature; a correlation-weighted mean was also computed and behaves almost
identically. Everything else is unchanged: same windows, same asymmetric
filter applied to the curvature series as to every other, $F_1$ at the 90th
percentile, block bootstrap, all three labellings.

Orientation was measured rather than assumed. Mean curvature is $0.711$ in
labelled crisis windows against $0.678$ in calm ones, so it rises under stress,
which is the direction the original work reports.

\begin{center}
\begin{tabular}{llccc}
\hline
Sample & Labels & $F_1$ TSI & $F_1$ curvature & $\Delta F_1$ ($p$) \\
\hline
OOS 2016--2026 & 16 episodes & 0.434 & 0.301 & $+0.126$ (0.011) \\
OOS 2016--2026 & 23 windows  & 0.447 & 0.337 & $+0.087$ (0.007) \\
OOS 2016--2026 & 25 episodes & 0.451 & 0.321 & $+0.101$ (0.003) \\
FULL 2000--2026 & 16 episodes & 0.340 & 0.335 & $+0.023$ (0.388) \\
FULL 2000--2026 & 23 windows  & 0.389 & 0.317 & $+0.074$ (0.018) \\
FULL 2000--2026 & 25 episodes & 0.386 & 0.316 & $+0.073$ (0.023) \\
\hline
\end{tabular}
\end{center}

Curvature lands where a reader familiar with this literature would expect it to
land: clearly above the Absorption Ratio, which scores 0.13 to 0.17 on the same
windows, and somewhat below the effective rank, which scores 0.34 to 0.38. It
is a mid-strength baseline rather than a weak one, and the honest summary is
that TSI is ahead of it in all six cells but not by the margin it holds
over the Absorption Ratio. Section~\ref{sec:multiplicity} reports what survives
once this comparison is folded into the corrected family, and only one of the
five nominally significant cells does.

\subsection{Correcting for Multiple Comparisons}
\label{sec:multiplicity}

This paper runs many related tests, and reporting each in isolation would
overstate the evidence. We therefore define the family in advance and correct
over all of it. Hypothesis~H1 is tested by comparing TSI against every
\emph{distinct} baseline (the effective rank, which is identical to the Vendi
score on a correlation matrix and so is counted once; the Absorption Ratio; and
the effective rank computed on the return matrix), on both samples, under all
three crisis labellings. With the Ollivier--Ricci curvature of
Section~\ref{sec:ricci} included, that is $4 \times 2 \times 3 = 24$ tests, and
all twenty-four enter the family, including those that were never going to be reported
as wins: choosing which tests belong is precisely the manoeuvre the correction
exists to prevent. We apply Holm--Bonferroni at a family-wise error rate of
$0.05$, which is uniformly more powerful than plain Bonferroni and assumes no
independence between tests.

\textbf{Sixteen of the twenty-four are significant before the correction and
nine after it.} The nine that survive are the eight wins over the Absorption
Ratio and over the effective rank on returns, under the 23-window and
25-episode labellings, on both samples, all with bootstrap $p$ below $0.0001$
against Holm thresholds between $0.0028$ and $0.0045$; plus one of the six
comparisons against curvature. The six comparisons against the
effective rank were ties before the correction and remain ties after it, so the
central finding of this paper is untouched: a correction cannot turn a tie into
anything else.

\textbf{Three results are lost, and all three sit on the same labelling.} The
win over the effective rank on returns out of sample under the 16-episode list
($\Delta F_1 = +0.179$, $p = 0.0060$ against a threshold of $0.0050$), and the
wins over the Absorption Ratio and the effective rank on returns over the full
history under the same list ($p = 0.040$ and $p = 0.044$). This is the
narrowest of the three labellings and the one with the least power, and the
paper already identified it as the weakest support for the Absorption Ratio
result before the correction was applied. The correction did not reveal a new
weakness; it quantified one that was already declared.

The practical reading is that the advantage over the Absorption Ratio is real
under the two broader labellings and does not reach family-wise significance
under the narrowest. We state the claim at that strength and no higher.

\subsection{Lead--Lag Analysis: a Coincident Index, Not a Forecast}
\label{sec:leadlag}

The F1-at-90th-percentile criterion used throughout asks whether TSI is
\emph{elevated during} documented stress windows. It is a contemporaneous
classification test, and it says nothing about whether TSI rises
\emph{before} stress. Because the framing of a ``fragility index'' invites the
stronger reading, we test the stronger claim directly.

We cross-correlate TSI against the OFR Financial Stress
Index~\cite{monin2019} at lags
$k \in [-15, +30]$ observations (each step $\approx$ 3 business days), where
$k > 0$ means TSI leads. Results on the full history ($n = 2{,}232$):

\begin{center}
\begin{tabular}{lccc}
\toprule
Signal & Peak $|\rho|$ at & $\rho$ at $k=0$ & Reading \\
\midrule
TSI raw & $k = 0$ & $+0.203$ & exactly coincident \\
TSI with memory & $k = -2$ & $+0.195$ & coincident (marginally lagging) \\
Absorption Ratio & $k = -15$ & $+0.063$ & lagging \\
\bottomrule
\end{tabular}
\end{center}

\begin{figure}[ht]
\centering
\includegraphics[width=\textwidth]{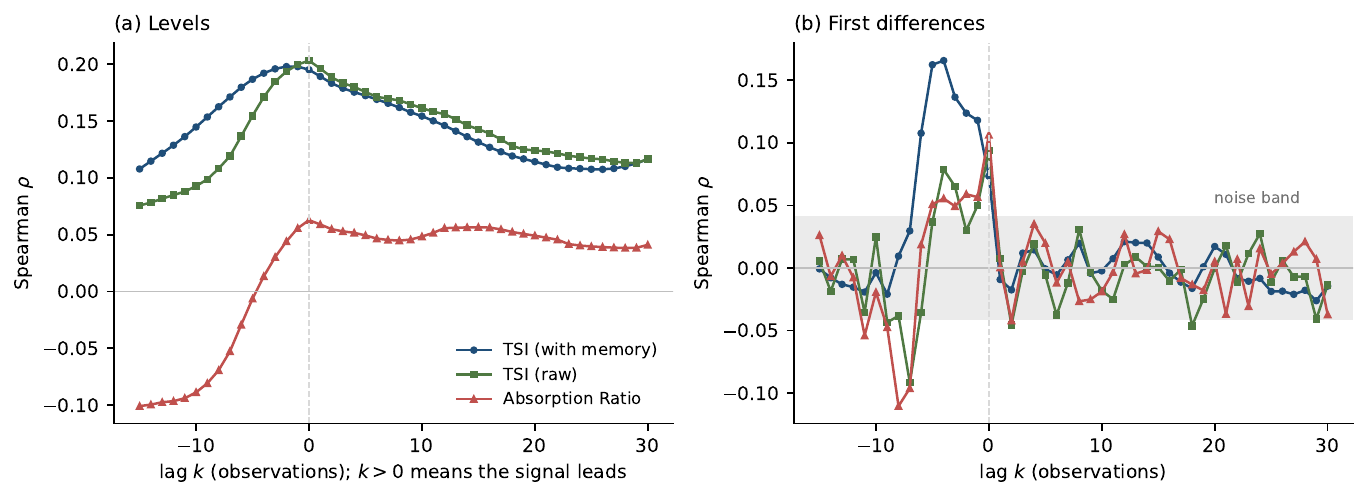}
\caption{Cross-correlation against the OFR Financial Stress Index. TSI with its
persistence filter is in dark blue, raw TSI in green, and the Absorption Ratio
in red. \textbf{(a)} In levels, TSI peaks at or just before zero lag and decays
symmetrically. \textbf{(b)} In first differences the same conclusion sharpens:
at positive lags, where a leading indicator would show its peak, the
correlation lies inside the noise band. The shape identifies a coincident state
index rather than a forecast.}
\label{fig:leadlag}
\end{figure}

Both series are strongly autocorrelated, so the absolute $\rho$ values in
levels are not the object of interest; the informative feature is the
\emph{shape} of the cross-correlation function, which decays symmetrically
about $k \approx 0$, the signature of a contemporaneous relationship rather
than a leading one. Repeating the analysis on first differences sharpens the
same conclusion: $\Delta\mathrm{TSI}$ raw peaks exactly at $k = 0$ ($\rho = +0.094$),
and at $k = +1$, with TSI leading by a single step, the correlation is
statistically indistinguishable from zero ($\rho = -0.009$ and $+0.008$ for the
memory and raw forms respectively, against a significance threshold of
$\pm 0.041$ at $n = 2{,}231$). The memory filter, being an exponential moving
average, introduces a small lag rather than an anticipation, peaking at
$k = -4$.

\textbf{What this does and does not change.} It does not affect the paper's
principal result: the F1 comparison was always a contemporaneous test, it
reproduces, and TSI outperforms the Absorption Ratio at every lag examined
($\rho \approx 0.20$ versus $0.06$). What it does change is the permissible
framing. TSI should be read as a \emph{state index}, closer in spirit
to Convective Available Potential Energy in meteorology, which reports that the
atmosphere is presently primed for severe weather without predicting when or
where a storm will form, than as an early-warning signal. A high reading
says the correlation network is presently in a concentrated, stressed
configuration. It does not say that stress is coming. We adopt that language
throughout and recommend that practitioners do the same.

\subsection{Four Original Scenarios}

\textbf{Bear Stearns (March 2008).} Raw TSI peaks \emph{before} the
rescue and falls afterward, while the Absorption Ratio keeps rising; TSI
with memory tracks the AR's sustained trajectory.

\textbf{Lehman Brothers (September 2008).} Raw TSI jumps sharply
within days of the collapse, a much sharper reaction than the AR,
and TSI with memory preserves this signal through October.

\textbf{European debt crisis (2011): the original failure case.}
Raw TSI falls to 0.17--0.30 during the crisis on the 10-bank network
while the AR stays elevated; TSI with memory corrects this.
Section~\ref{sec:large_network} shows this failure does not replicate on a
larger, diversified network.

\textbf{AI sector: the 2025 tariff shock.} The same failure-and-correction
pattern recurs on the tech-stock network ($n=11$).

\subsection{Large-Network Robustness and the Central Finding}
\label{sec:large_network}

Running the full pipeline on a 42-asset network diversified across seven
sectors, 2006--2026, and normalising to z-scores:

\begin{table}[ht]
\centering
\caption{Ending z-scores by network composition.}
\label{tab:zscores}
\begin{tabular}{lrrrr}
\toprule
 & Financials & Tech & Diverse & Full \\
 & ($n=10$, epicentre) & ($n=8$, non-ep.) & ($n=9$, mixed) & ($n=42$) \\
\midrule
Bear Stearns & $-0.28$ & $-0.42$ & $-0.26$ & $-0.12$ \\
Lehman       & $0.10$  & $1.20$  & $1.27$  & $\mathbf{2.61}$ \\
Euro debt    & $-1.09$ & $2.18$  & $\mathbf{5.35}$ & $2.13$ \\
\bottomrule
\end{tabular}
\end{table}

The pattern sits in the epicentre column. During the European debt crisis the
ten-bank basket, which \emph{is} the epicentre sector, ends at $z = -1.09$,
below its own historical mean while the crisis is still running, whereas the
diverse and full baskets end at $+5.35$ and $+2.13$; the same ordering holds at
Lehman. That collapse of the reading during the episode is the forgetting
behaviour the memory filter was built to repair, and it appears only where the
basket is concentrated on the epicentre.

Node count alone does not explain it: the 8-node tech-only basket does
\textbf{not} exhibit the forgetting pattern. \textbf{The variable that
matters is whether the tested basket is concentrated in the sector that is
the actual epicentre of the crisis.} Any basket that avoids that
concentration, whether small-and-diverse or large-and-diverse, retains
an elevated reading.

Replication in COVID (z = 1.35, Feb--Apr 2020) and the 2022 rate-hike
selloff (z = 1.41, Jan--Oct 2022) confirms the pattern in two independent
episodes.

\subsection{Cross-Market Test: Crypto, Commodities, FX, and Sovereign Debt}

\textbf{Two clear successes.} The all-time peak in the crypto series
(z = 3.9) falls in May--June 2022, the Terra/Luna collapse, with
MATIC-USD flagged as the most-connected asset. The all-time peak in
sovereign debt (z = 3.6, August--December 2023) coincides with the historic
bond selloff, with MBB (the highest-duration instrument) flagged as
responsible.

\textbf{Two episodes outside what TSI is built to measure.} The
September 2022 pound sterling / UK gilt crisis: TSI with memory stayed
negative throughout, because the crisis was largely contained to
sterling-linked instruments rather than reorganising the broader FX
correlation structure. The 2022 Ukraine invasion: a $\sim$3-month lag before
TSI rose, consistent with geopolitical shocks propagating through
price co-movement over weeks rather than instantaneously.

These are not detection errors but \emph{scope boundaries}: TSI detects
correlation-structure breaks (episodes where normally semi-independent assets
suddenly co-move), not directional price events. This yields a precise,
falsifiable scope claim.


\subsection{Attempted Replication on Shorter Series}

We attempted to repeat the calibration on the bank network (train:
Bear Stearns + Lehman 2008; test: European debt crisis 2011) and the AI
network. Both failed to reproduce the OFR result, and not because the filter
stopped working, but because the highest TSI with memory readings in
the training period systematically fell \emph{outside} our defined crisis
windows (they clustered on real but undocumented events).

\textbf{Finding.} With short series (5 years, 2--3 labelled episodes), the
definition of the ground-truth event list dominates the calibration result
far more than the choice of hyperparameters. The 26-year, 20+-episode OFR
series has enough independent events to support reliable out-of-sample
calibration; 5-year, 2--3-episode series do not.

\subsection{Persistent Homology}

The correlation-network H1 construction showed no discriminative pattern
(mean 0.03--0.07 across all scenarios; calm and crisis periods
indistinguishable). We attribute this to insufficient topological richness
in an 8--11-node graph.

The Takens-embedding construction tells a different story. The all-time
maximum (30 March 2020, COVID) sits $\sim$12 standard deviations above the
historical mean (0.104 $\pm$ 0.110). \textbf{Fifteen of the twenty highest
readings fall inside documented crises}, including October--November 2011,
exactly the episode the small-network raw TSI missed.

\subsection{The signed channel, and a head-to-head against the balance index}
\label{sec:signed}

Three questions follow from the previous paragraph, and we answer them on the
same windows, with code in the repository.

\textbf{First: does the sign of the triangle carry independent information?
No.} Replacing $\operatorname{Tr}(A^3)$ by $\operatorname{Tr}(\rho^3)$, which
counts balanced minus frustrated triangles, changes nothing: the two series
correlate at $1.000$ across all 2{,}232 windows, and the resulting $F_1$ moves
by $0.000$ (95\% CI $[-0.005, +0.005]$). The reason is structural rather than
particular to this data. Positive semi-definiteness bounds how frustrated a
triangle in a correlation matrix can be, so signed and unsigned triangle counts
are nearly collinear by construction; the measured ratio averages $0.92$. The
sign channel that the balance literature exploits is not reachable through the
triangle count alone, and we retire this extension rather than carry it
forward.

\textbf{Second: the balance index as an attribution rule.} The local balance of
Bartesaghi et al.\ shares with $\mathrm{diag}(A^3)$ the property of having no
$k$ to select, so it belongs in the comparison above on merit. On the
same benchmark it scores $0.329$, $0.552$, $0.724$ and $0.841$ for one to four
epicentres, against $0.992$, $0.996$, $0.972$ and $0.966$ for the triangle
rule, with chance at $0.188$ to $0.750$. Because the generator contains no sign
structure, which is precisely what a balance index is built to read, we
repeated it with half of each stressed block loading negatively on its factor,
so that frustrated triangles are present by construction. The ordering does not
move: $0.992$, $0.988$, $0.979$, $0.968$ against $0.350$, $0.567$, $0.719$,
$0.838$.

\textbf{Third: detection, where we decline to claim a win.} On our windows the
balance index scores poorly, but the reason disqualifies the comparison rather
than settling it. The index \emph{saturates}: with 20-day windows over eight
nodes it sits at its maximum of $1$ on 26\% of windows in its weighted form and
70\% in the binary form the same authors report, because short-window
correlation matrices are almost entirely positive and an all-positive network
is perfectly balanced. On the 42-asset network of
Section~\ref{sec:large_network} lengthening the window makes it worse rather
than better, the fraction of windows at $\kappa = 1$ rising from 10\% at 45 days
to 24\% at 100 and 63\% at 400.

That last observation, however, is a statement about small networks and not
about the index, and the distinction decides who the finding is about. Repeating the measurement on 447 S\&P~500 constituents, and again on
464 over a ten-year panel, the saturation disappears: $\kappa$ ranges over
almost the whole unit interval, with a median of $0.44$ on the first panel and
$0.03$ on the second at a 60-day window, and at most $2.0\%$ of windows sit at
the maximum at any window length from 45 to 500 days. The measurement is
reproduced by \texttt{code/validation/balance\_saturation\_sp500.py}. The reason is combinatorial. The number of triangles
grows as $n^3$, so in a large network a modest fraction of negative correlations
already produces enough frustration to move the index, whereas a network of a
few dozen nodes is close to balanced whatever the market is doing. Saturation is
therefore a property of \emph{small} correlation networks, and the regime the
original authors work in, 199 to 385 assets, is not one of them.

Two consequences follow, and neither favours us. Our own networks of eight and
42 nodes are exactly the regime where their index cannot be evaluated, so no
ranking produced on them would be informative about it, and none is given here. And the
level of $\kappa$ is not portable: the median moves from $0.44$ to $0.03$
between two panels of the same market a few years apart, so a fixed threshold
calibrated on one dataset need not fire at all on another. What we do
report is a check in the other direction: on our data their headline
association reproduces, with the Spearman correlation between the balance index
and mean absolute correlation at $0.94$, $0.89$ and $0.71$ for the three window
lengths against the $0.877$ they report, and with the conditional mean of
forward average returns decreasing monotonically across balance quintiles at
45-day windows, as they describe.

\textbf{A methodological note that this exercise forced, and that applies to
the like-for-like table of Section~\ref{sec:spectral}.} $F_1$ at a fixed percentile is a fixed alarm
budget only if the metric takes distinct values near the threshold. A saturating
metric flags far more than the intended fraction of windows and scores an
inflated $F_1$ through recall alone; the binary balance index flags 70\% of
windows at its 90th percentile and reaches $F_1 = 0.537$ with a precision of
$0.425$ against a base rate of $0.406$, which is no better than chance. We
verified that no benchmark reported in this paper has this property. Every
series in that table, raw and filtered, takes distinct values throughout and
flags exactly 10.0\% of windows at its 90th percentile; the check is printed by
\texttt{code/validation/samefilter\_benchmarks.py} in the repository.

\section{Discussion}

\textbf{What the index is worth, stated as a package.} No single comparison in
this paper is decisive on its own, and we do not present one as such. Taken
together they describe something narrower and, we think, more useful than a
win. On detection, TSI matches the best spectral measure available: a tie
in all six labelling-by-sample cells, with the point estimate in TSI's
favour in five of them, which is a consistent direction that never separates
from zero at this sample size. At the same alarm budget its alarms are the
cleanest of any metric tested, 4.0\% unexplained against 14.7\% for the
effective rank and roughly 59\% for the Absorption Ratio, though that is the
same comparison re-expressed and carries the same verdict. It reaches that
level from a different primitive, network topology rather than variance
decomposition, sharing only half its alarms with its nearest relative. And it
arrives with a node-attribution layer that requires no selection rule, ties the
best adaptive spectral rule, and cannot be misconfigured the way a fixed or
badly chosen $k$ can be.

None of those four is a headline on its own. The defensible claim is the
conjunction: \emph{equal detection performance, cleaner alarms in point
estimate, a different construction, and attribution included at no extra
parameter}. A practitioner already running an effective-rank monitor would not
switch on the strength of the detection numbers alone, and we do not suggest
they should. What they would gain is the attribution layer, which their current
tool supplies only after they commit to a component count and get it right.

The central methodological lesson is that \textbf{whether a tested basket is
concentrated
in the sector at the epicentre of a given crisis, not raw network size,
determines whether raw TSI needs a memory correction.} On a network
concentrated in the epicentre sector (10 banks during banking crises), raw
TSI behaves as a sharp but forgetful alarm. On a network that avoids
that concentration, whether large-and-diverse (42 assets) or
small-but-diverse (9 assets spanning seven sectors), raw TSI already
exhibits sustained-elevation behaviour, replicated across three independent
crises.


\section{Limitations}
\label{sec:limits}

\paragraph{On real matrices the attribution reduces to the degree.}
The per-node layer is defended in this paper on a synthetic benchmark where the
ground truth is known, and there $\mathrm{diag}(A^3)$ scores 0.966 to 0.996
against 0.933 to 0.961 for the far simpler node degree, $\sum_j |\rho_{ij}|$.
That gap does not survive contact with real data. Across 245 windows of 464
S\&P~500 constituents, the cross-sectional rank correlation between
$\mathrm{diag}(A^3)$ and the degree averages \textbf{0.996}, and a cubic in the
degree explains $R^2 = 0.992$ of the triangle count. We then asked whether the
remaining 0.8\% carries anything of its own, since that residual is the only
part of the quantity that is not a restatement of the degree. It does not: its
rank correlation with forward five-day returns is $+0.002$ with a bootstrap
interval spanning zero, and its apparent association with forward volatility
($+0.187$) disappears once current volatility is partialled out ($+0.016$,
interval $[-0.003, +0.035]$), while the degree itself retains a small but
non-zero effect there. The practical reading is that on correlation networks of
real assets, naming the culprit with $\mathrm{diag}(A^3)$ and naming it with the
degree give the same answer, and a practitioner should use the degree, which
costs one matrix row-sum. What the triangle count buys is what
Section~\ref{sec:attrib} claims and no more: no parameter to misspecify, and a
measured advantage in a regime where the ground truth can be checked.

\begin{itemize}
\item The large-network result (Section~\ref{sec:large_network}) is a single test on one 42-asset
  universe; it has not been repeated with different sector compositions or
  other crises.
\item The attempted short-series recalibration is a negative methodological
  result; we did not re-derive ``correct'' crisis windows to avoid
  indefinitely re-tuning ground truth.
\item The Takens-embedding H1 result uses one embedding dimension and delay,
  chosen once; a proper replication would test robustness to these choices.
\item The statistical significance test evaluates the memory filter against
  the exact objective it was designed for; it is not independent evidence of
  general superiority.
\item Three different ground-truth lists are in use across the paper
  (Section~\ref{sec:oos_results}). Every F1 figure uses one of them
  consistently, but the precision, recall and unexplained-alarm figures do not
  come from the same list as the F1 figures, and no result should be quoted as
  if they did.
\item The attribution result of Section~\ref{sec:attrib} is established on
  synthetic networks. Against a spectral comparator with an adaptive $k$ it is
  a tie, not an advantage; the advantage claimed is the absence of a selection
  rule, not accuracy.
\item The win over the Absorption Ratio does not survive the narrowest of the
  three crisis lists (Section~\ref{sec:spectral}), and it is one of the three
  results that a Holm--Bonferroni correction over the family of eighteen tests
  removes (Section~\ref{sec:multiplicity}). The tie against the effective rank
  survives all three lists and is unaffected by the correction, as a tie must
  be.
\item Nothing in this document constitutes investment advice.
\end{itemize}


\section{Where This Stands}

Several questions that a reader would reasonably raise have already been
answered, and it is more useful to say so than to list them as future work.

\textbf{Settled, and reported above.} Whether the conclusions survive a
correction for multiple comparisons: Section~\ref{sec:multiplicity} defines the
family of twenty-four tests and applies Holm--Bonferroni, and reports what does
not survive it. Whether the index beats the nearest structural precedent,
Ollivier--Ricci curvature: Section~\ref{sec:ricci} runs that head-to-head, which
earlier versions declared and did not perform. Whether declining to rank the
global balance index on small networks was concealing a loss: it was not, and
Section~\ref{sec:signed} reports the tie on a 464-asset universe where the index
is well behaved, under the original authors' own definition of a systemic
event. Whether the attribution advantage
survives a spectral rule that chooses its own number of components: it does
not, and the claim was reduced accordingly to robustness rather than
resolution. Whether the alarm-quality figure holds for the competitors too:
computed for all of them. Whether the results depend on which list of crisis
episodes is used: all three are reported. Whether signing the correlation
recovers the single-asset blind spot: it does not, and the reason is
structural. Whether the per-node layer sees something the node degree does not:
on real matrices it does not, and Section~\ref{sec:limits} says so with the
numbers.

\textbf{Open, and worth doing.} A systematic sweep of the
Takens embedding parameters, which would not change any conclusion here. And an
attribution test on real data, which needs
a partial ground truth, for instance from regulatory post-mortems of documented
episodes; we note that after the degree result of Section~\ref{sec:limits} the
expected outcome of that test is fairly clear, which lowers its value.

\textbf{Not attempted here, deliberately.} Everything in this paper is a
backtest of a state reading. Turning that into an allocation rule is a
different exercise with a different evidential standard, and mixing the two
would weaken both.

Issues, pull requests and replication attempts are welcome at
\url{https://github.com/BiomeMakers/TSI-OmegaS}. For production licensing:
\texttt{acedo@biomemakers.com}.

\section{Conclusion}

TSI's apparent need for a memory correction turns out to be largely a
property of small, sector-homogeneous, or crisis-epicentre-concentrated
networks, not of TSI itself. On a large, diversified network, raw
TSI already tracks sustained stress correctly. The memory filter remains
a validated, statistically significant, out-of-sample-tested fix for the
small or concentrated-network case, which is the realistic case for many
practical applications (a single sector, a single portfolio, a single
institution's counterparties).

Across all tests in this paper, namely four original scenarios, out-of-sample
calibration, statistical significance testing, the like-for-like spectral
benchmark under three crisis labellings, the adaptive-$k$ attribution control,
the lead--lag analysis, the disentangled large-network robustness check and the
persistent-homology extension, the evidence supports a specific and bounded
conclusion. TSI matches a properly sharpened spectral measure at
detecting stress rather than beating it. It equals one,
from a different construction, with the cleanest alarms of anything we tested,
and it brings with it an attribution of each reading to a specific asset that
needs no component count to be chosen and therefore cannot be chosen wrongly.
That combination is what we would ask a practitioner to weigh, complementary to
and not a replacement for established indices such as the Absorption Ratio.

\bibliographystyle{unsrt}

\vspace{1em}
\noindent\textit{Correspondence:} Alberto Acedo, Biome Makers Inc.
(\texttt{acedo@biomemakers.com}).
The author declares no competing financial interests beyond patent
applications filed by Biome Makers Inc.\ relating to the index described
here.

\end{document}